\def\lsmo{La$_{1.2}$Sr$_{1.8}$Mn$_2$O$_7$ }
\def\lsmox{La$_{2-2x}$Sr$_{1+2x}$Mn$_2$O$_7$ }
\documentstyle[prl,aps,amsfonts,amssymb,floats,epsfig,twocolumn]{revtex}
\begin{document}

\draft
\twocolumn[\hsize\textwidth\columnwidth\hsize\csname
@twocolumnfalse\endcsname
\title{Work function changes in the double layered manganite \lsmo}
\author{K. Schulte, M. A. James, L. H. Tjeng, P. G. Steeneken, G. A. Sawatzky }
\address{Material Science Center, University
 of Groningen, Nijenborgh 4, 9747 AG Groningen, The Netherlands}
\author{R. Suryanarayanan, G. Dhalenne, A. Revcolevschi}
\address{Lab. de Chimie des Solides, B\^{a}t. 414, CNRS, UA 446,
Universit\'{e} Paris-Sud, 91405 Orsay, France}
\date{\today}
\maketitle
\begin{abstract}
We have investigated the behaviour of the work function of \lsmo
as a function of temperature by means of photoemission. We found a
decrease of 55 $\pm$ 10 meV in going from 60 K to just above the
Curie temperature (125 K) of the sample. Above $T_C$ the work
function appears to be roughly constant. Our results are exactly
opposite to the work function changes calculated from the
double-exchange model\cite{furukawa}, but are consistent with
other data. The disagreement with double-exchange can be explained
using a general thermodynamic relation valid for second order
transitions\cite{marel} and including the extra processes involved
in the manganites besides double-exchange interaction.
\end{abstract}
%
\vskip2pc]
\narrowtext
\section{INTRODUCTION}
Doped manganese oxides containing a significant proportion of
Mn$^{4+}$ ions exhibit colossal magnetoresistance (CMR), an effect
whereby the resistivity around the Curie temperature ($T_C$) of
the material dramatically diminishes when a magnetic field is
applied. This effect was first discovered in the cubic perovskite
manganese oxides\cite{kusters} and later extended to other
manganites from the same family, containing one or more MnO$_2$
layers\cite{mahesh,moritomo}. Within this so called
Ruddlesden-Popper series, the double layered (n=2) branch has
received tremendous interest as the CMR effect is even stronger
than in the 3D perovskites (n=$\infty$) compounds, reaching MR
ratios of 3000\% at 1 Tesla, compared to approximately 110\% for
the cubic compounds\cite{moritomo}. The strong anisotropy in these
layered materials, rendering the materials quasi two-dimensional,
is responsible for this increased effect. Another important
discovery was made with poly-crystalline samples or thin films
containing grain boundaries (GB). Here the CMR effect was found to
extend over a much wider temperature range below $T_C$ and set in
already at smaller applied fields (B < 1T), which is advantageous
in applications\cite{hwang}. A thorough investigation into the
precise influence of grain boundaries on CMR was then
started\cite{klein} and, for instance, Klein {\sl et al.} explain
their findings by assuming that the GB region is structurally
disordered and the accompanying stress fields strongly reduce the
local Curie temperature. This means that below the bulk $T_C$, the
GB region still remains in the high temperature paramagnetic (PM)
phase whereas the grains themselves are now ferromagnetic (FM). A
substantial difference in work function between the PM and FM
regions, and the subsequent build up of a charge depleted region
around the GB, can then explain the increased resistivity, the
strongly non-linear IV curves, and the magnetic field dependence
of the resistivity below $T_C$, of films containing one or more
grain boundaries, in comparison to thin films without GB's. This
difference in work function between the PM and FM phases arises
naturally from the double-exchange (DE) model\cite{zener}, which
is the traditional starting point for explaining the MR phenomenon
in the manganites, and in the following section we will take a
more in-depth look at this model. Although the simplified DE model
alone can by no means fully account for the rich and diverse
properties of the actual compounds, it is widely accepted as a
good basis for their explanation. In real materials, other aspects
play a role, such as the influence of the competing
antiferromagnetic super exchange interaction\cite{perring}, or the
strong interplay between electrons and lattice\cite{millis}, to
name but a few. However, in view of the experiments by Klein {et
al.} it remains essential to investigate whether this work
function difference predicted by the DE model does indeed exist.

\section{The double-exchange model}
In this section we will review parts of the DE model, in order to
present a complete picture. In Fig.\ 1 we qualitatively
demonstrate the idea of this form of direct exchange.
\begin{figure}[ht]
\centerline{\epsfig{figure=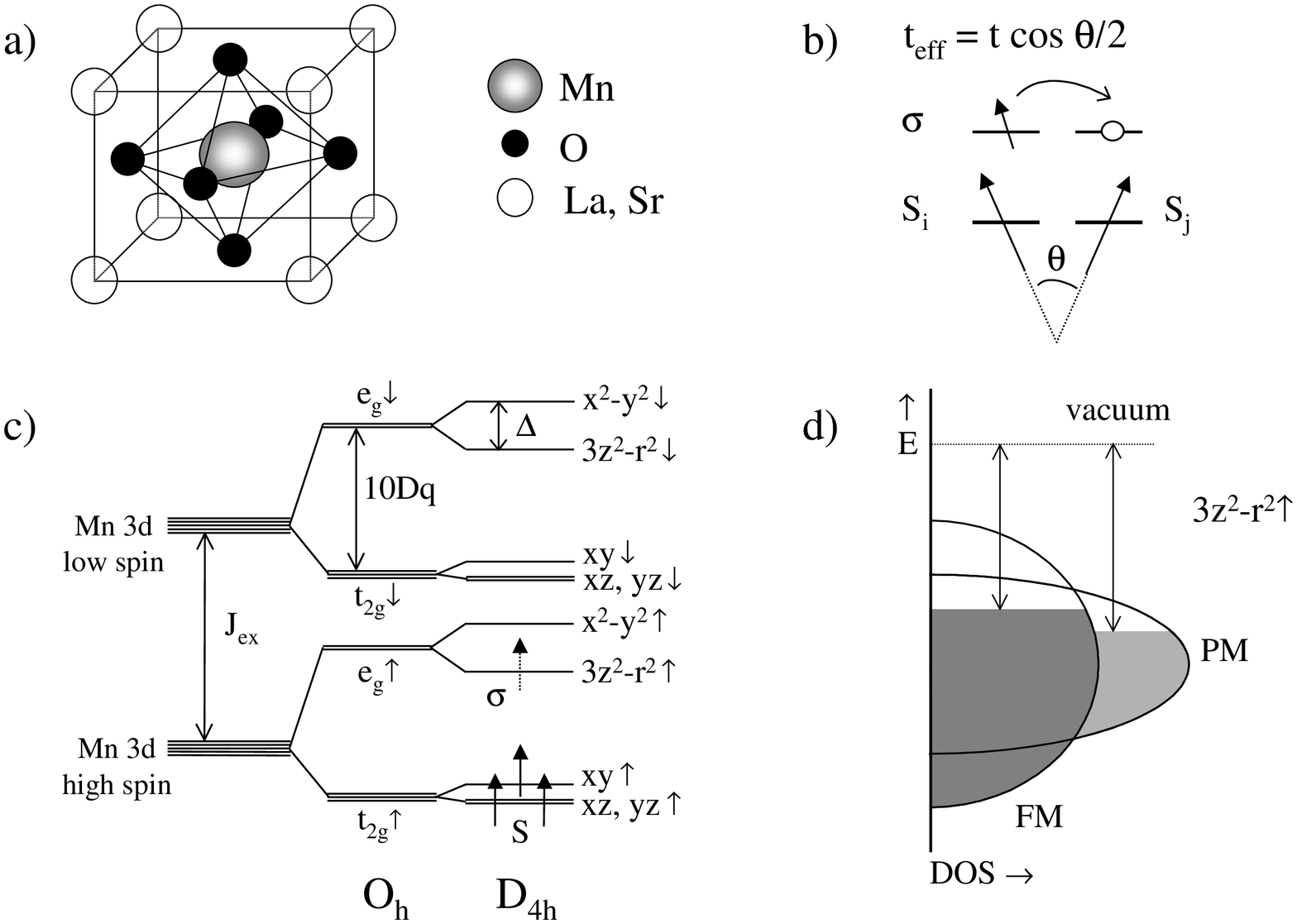,width=7.4cm,angle=270,clip=}}
\caption{a) Octahedral surroundings of a Mn ion. In
La$_{1.2}$Sr$_{1.8}$Mn$_2$O$_7$ the symmetry is further reduced to
D$_{4h}$, which corresponds to stretching the cube in the z
direction. b) Definition of the parameters involved in the DE
model. c) Full 3d level splitting for a Mn ion in D$_{4h}$. d) One
electron conduction band picture of the effects of DE interaction
in the PM and FM phase.} \label{bands}
\end{figure}
According to Hund's rules, each separate Mn$^{3+}$ or Mn$^{4+}$
ion will be in high spin configuration, and this will be our
starting point. In fact, the DE model assumes that the
intra-atomic exchange integral $J_{e\!x}$ can be considered
infinite. When placed in octahedral surroundings (Fig.\ 1a) the Mn
$3d$ orbital splits up further into a $t_{2g}$ and an $e_g$ level.
In the layered materials the symmetry is further reduced to
tetragonal, because of the slight difference in Mn-O distances
within the MnO$_2$ bilayer plane compared to those directed out of
it\cite{mitchell,medarde}. In Fig.\ 1c this complete $3d$ orbital
splitting is drawn, and also the localized $t_{2g}$ spin $S=3/2$
and the itinerant $e_g$ spin $\sigma$ are indicated. Although in
$D_{4h}$ symmetry the $t_{2g}$ level theoretically splits up into
a $b_{2g}$ ($xy$) and an $e_g$ ($xz,yz$) level, $O\: 1s$ XAS
measurements have indicated that a splitting between these two
levels was not clearly observed and thus the $t_{2g}$ level
remains basically unchanged\cite{park}. In the same measurement an
effective splitting of 0.4 eV {\sl was} observed between the two
$e_g$ orbitals, which has turned out to be very important in
understanding our measurements, as will become clear later on.\\
The lowest unoccupied level for a Mn$^{4+}$ ion is now the $3z^2
-r^2$ majority spin orbital, which can contain at maximum one
electron. At this point we would like to stress that this level
diagram is valid for {\sl on-site} excitations only. If we want to
look at electron hopping between adjacent Mn ions, the energy
needed for transfer will also depend on the number of initial $3d$
electrons on both ions. In the DE model larger charge fluctuations
are neglected: electron transfer from a $3+$ to $3+$, or from a
$4+$ to $4+$ ion is costlier in energy by a factor $U$, the
on-site Coulomb repulsion energy, than hopping from a $3+$ to a
$4+$ ion. Therefore, these processes are not taken into
consideration in the double-exchange model, effectively this means
that one assumes $U$ to be infinite. Furthermore the assumption
$J_{e\!x}\rightarrow\infty$ excludes the low spin configuration of
$\sigma$ and $S$ and will also not allow electron transfer to an
ion with $t_{2g}$ spin $S$ anti-parallel to the initial one. These
assumptions concerning $U$ and $J$, combined with the crystal
field splitting, lead again to a description of the charge
carriers in terms of a single electron, band picture. The only
ingredient remaining is the spin-correlation between adjacent
manganese ions. This is reflected in the dependence of the
effective hopping integral $t_{ef\!f}$ on the angle $\theta$
between neighbouring $t_{2g}$ spins $S_i$ and $S_j$ (see Fig. 1b)
via: $t_{ef\!f} = t\cos\, (\theta/2)$. In the FM phase, $\theta =
0$, giving $t_{ef\!f}=t$. In the PM phase the average angle will
be 90$^\circ$ and this will reduce $t_{ef\!f}$ to $t/\sqrt{2}$.
This difference of a factor of $\sqrt{2}$ in $t_{ef\!f}$ is the
maximum effect that can be expected, because it uses the
assumption that the intra-atomic exchange splitting J$_{e\!x}$ is
infinite, or at least much larger than the bandwidth W of the
conduction band. In this case, since W is directly proportional to
$t_{ef\!f}$, the same change across $T_C$ is expected for W\@. In
this simple band picture, if we keep the center of the band fixed
throughout the transition, we see that (at fixed filling) the
chemical potential $\mu$ will have to change upon crossing $T_C$
(Fig.\ 1d). Or, in terms of the work function $\Phi$,
$\Phi_{F\!M}<\Phi_{P\!M}$ if the band is more than half filled and
{\sl vice versa} in the case of a less than half filled band. In
our sample, La$_{1.2}$Sr$_{1.8}$Mn$_2$O$_7$, the $3z^2-r^2$ level
on average holds 0.6 electron and the band is thus more than half
filled. This intuitive picture for the chemical potential change
has been confirmed, within the double-exchange model, by dynamical
mean field calculations by Furukawa\cite{furukawa,comment}. He
also predicts that the change in $\mu$ or $\Phi$ will be in the
order of 10\% of the bandwidth, meaning roughly 0.1 eV\@. This is
a rather significant change which should in principle be
observable by angle resolved photoemission measurements and would
definitely be large enough to explain the aforementioned effects
observed on samples containing GB's. There is a problem however
with such a direct observation in the case of strongly correlated
systems: the photoemission spectrum will, in general, not just
consist of simple single electron-like peaks, but rather be made
up of a reduced intensity, quasi-particle peak, combined with an
incoherent background, or might even totally lack a quasi-particle
peak. Since a pseudo-gap has been observed in the manganites and
the intensity of the photoemission spectrum is strongly reduced in
the vicinity of the Fermi Energy E$_F$\cite{saitoh}, such a direct
observation of the chemical potential change in the manganites is
not possible. Nevertheless, photoemission can be used in an
alternative way, that does permit us to observe these changes, as
we will illustrate in the next section.

\section{experiment}
We used high quality \lsmox single crystals, with $x = 0.4$ Sr
doping, grown by the traveling solvent-floating zone
method\cite{sury}. $T_C$ of these samples is 125 K, and here a
sharp drop in both the $ab$ and $c$ axis resistivity of two orders
of magnitude is observed. We prepared the samples for measurements
by cleaving {\sl in situ} at 60 K by means of a top post, exposing
a clean and flat (001) surface. All measurements were performed
using an Omicron helium discharge lamp (photon energy 21.22 eV), a
VG Clam 2 electron analyzer (acceptance angle 8$^\circ$, overall
(uncorrected) energy resolution 50 meV) and a coolable Janis
cryostat. The temperature was measured using a calibrated
Pt-thermocouple. The pressure inside the vacuum chamber was always
$3\cdot10^{-11}$ mbar or better. In order to see the low kinetic
energy cutoff of the spectrum, which gives us the work function,
we have to bias the sample by a few volts, in order to push the
`zero' kinetic energy electrons out of the energy region where the
analyzer can no longer respond linearly to counts (roughly the
region from 0 to 1 eV). It is, however, important to keep the bias
as low as possible to avoid distortion of the spectra due to the
build up of electric fields.\\ We note here that this method can
not provide accurate {\sl absolute} values of the work function,
but as we are only interested in {\sl changes} in the work
function, this method is applicable, provided of course, that the
changes are large enough compared to the energy resolution and
statistics of the measurements. In order to determine the accuracy
of our system, we performed a cyclic temperature dependent test
measurement on a poly-crystalline silver sample (see Fig.\  2),
which was sputtered and annealed twice {\sl in situ}. Fig.\ 2a
shows an example of the Fermi edge and Fig.\ 2b shows the low
energy cutoff, both at 2 volt bias and T=50 K. In Fig.\ 2c we show
the work function as a function of temperature.
\begin{figure}[ht]
\centerline{\epsfig{figure=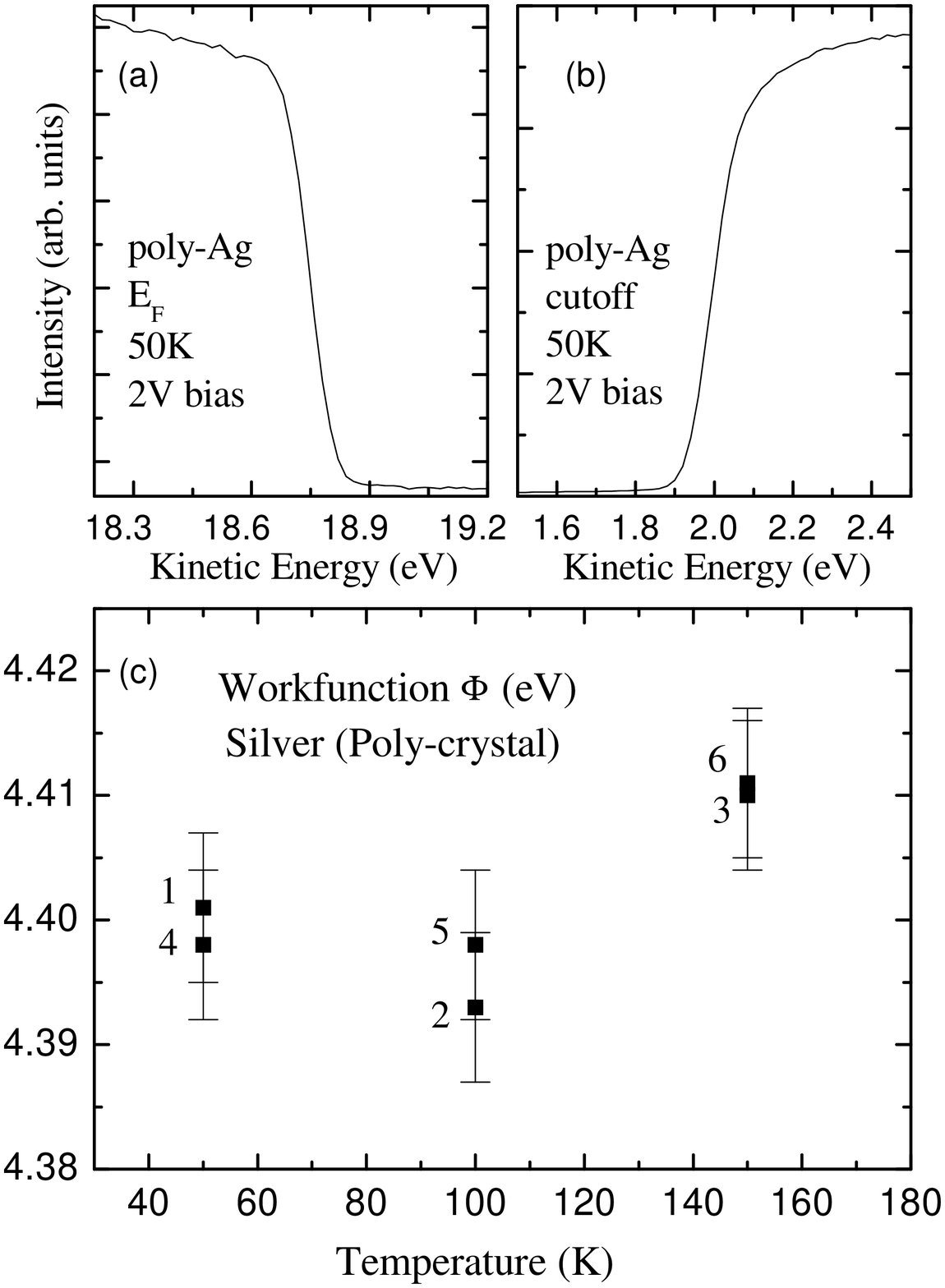,width=8.4cm,clip=}} \caption{a)
Ag Fermi level cutoff. b) Low kinetic energy cutoff of our biased
Ag sample. c) Results of the work function measurement on a
poly-crystalline Ag sample as a function of temperature and at 2
volt applied bias.} \label{ag}
\end{figure}
At each temperature the cutoff position was also determined as a
function of bias to monitor the optimum bias voltage range. We
assumed that the cutoff is defined by a step-function convoluted
with a Gaussian describing the broadening of the cutoff in energy,
due to the finite energy resolution of the analyzer and
temperature. We noticed, from fitting, that measurements with a
bias between 1.5 and 3 volt were trustworthy on all accounts. At
each measured temperature, the reproducibility is remarkable (less
than 5 meV difference), but over the entire temperature range of
50 K to 150 K, we observed an average value of 4.403 $\pm$ 0.013
eV, assuming the work function of silver should not show a
temperature dependence. The apparent contradiction between
reproducibility at a particular temperature and dissimilar values
at different temperatures can be due to slight movements of the
measuring spot with temperature, caused by expansion and
contraction of the cryostat. On a poly-crystalline sample this can
mean a difference in average value of the work function over the
various exposed crystal faces within the spot, and should
therefore not give any problems on a single crystal. We also
performed the silver measurements to obtain an accurate value for
the Fermi edge, since on \lsmo a clear edge is not observed in
ARPES measurements\cite{dessau}.
\\ The numerical value of the work function is obtained through
\begin{equation}
\Phi = 21.22 - \left.(E_F - CO)\right|_{\scriptstyle bias} +
{\displaystyle bias}.
\end{equation}
Where 21.22 (eV) stands for the energy of the incoming UV light.
The Fermi energy E$_F$, and the cutoff energy CO, are both
measured with a bias voltage applied.\\ Before discussing the
results it is important to stress that these very surface
sensitive measurements need to be performed quickly, because the
surface of \lsmo deteriorates rapidly\cite{saitoh}, probably
resulting in a more lanthanum rich compound at the
surface\cite{XPS}.

\section{Results}
In Fig.\  3 we show the results of the measurements on the \lsmo
single crystal. Fig.\ 3a shows an example of a low energy cutoff
at 2 volt bias and 60 K. In Fig.\ 3b we show three broad spectra,
one taken at the start of the measurements right after the cleave
(black, solid), one just after finishing the work function
measurements (grey, dashed) and the last one taken 9 hours after
the cleave (black, dotted) all at 60K.
\begin{figure}[ht]
\centerline{\epsfig{figure=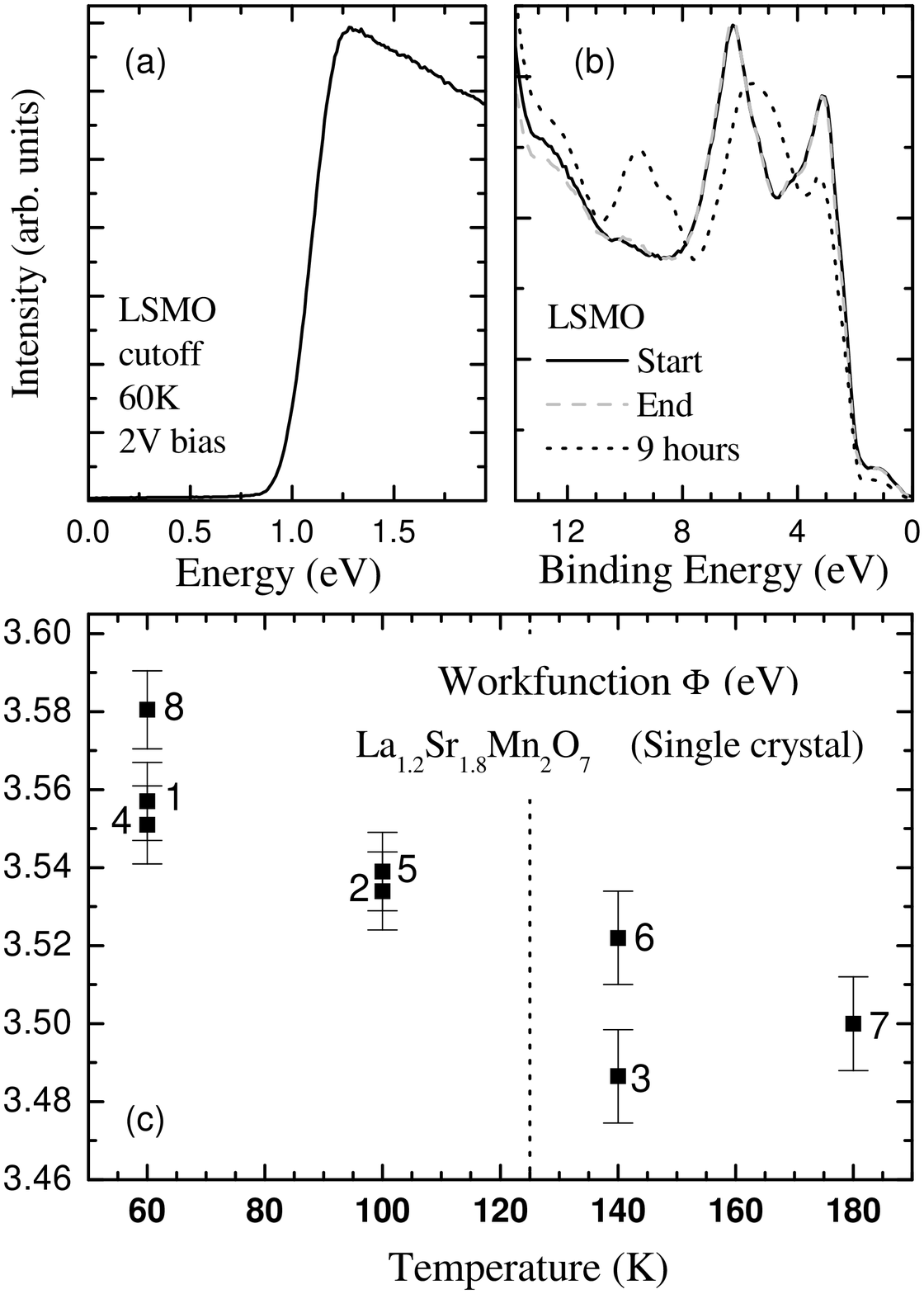,width=8.4cm,clip=}} \caption{
a) Low kinetic energy cutoff of our biased sample. b) Overview
photoemission spectra of La$_{1.2}$Sr$_{1.8}$Mn$_2$O$_7$
indicating the deterioration of the surface of the sample in time.
c) Results of the work function measurement on single crystal
La$_{1.2}$Sr$_{1.8}$Mn$_2$O$_7$ as a function of temperature and
at 2V applied bias (notice the difference in scale compared to the
Ag results).} \label{lasr}
\end{figure}
The spectrum taken just after the last work function measurement
shows an increase in the 9 eV peak, but this is still small
compared to the "fully aged" sample. After this point in time we
found the values of the work function increased as well and we
were not able anymore to reproduce quantitatively the values of
the work function measured within the first two hours after the
cleave. The decrease of the work function with temperature
remained also in the aged sample however. The bottom panel, Fig.\
3c, shows the work function as a function of temperature for
La$_{1.2}$Sr$_{1.8}$Mn$_2$O$_7$. Both temperature runs show a
decrease of the work function with increasing temperature below
$T_C$, and a roughly constant value of $\Phi$ above it. The Curie
temperature of 125K is indicated by a dotted line. The numbers
indicate the order in which the points were measured. The square
labeled 8 shows the increased work function that lead us to stop
the measurements. The downward trend in $\Phi$ with temperature
was found in all successful cleaves, but, since the deterioration
of the surface is immediately reflected in our surface sensitive
measurements, the presented data is from the measurement where we
were able to complete {\sl two} entire temperature cycles before
the surface had altered. The other recurrent behaviour we found in
different measurements is the increase of the work function with
time, i.e. with surface degradation. This is not surprising
considering the low starting value of the work function: $\pm$
3.56 eV at 60 K, which is almost a full eV lower than that of
silver.\\ Quantitatively we find $\Phi$ to decrease by $\approx 55
\pm 10$ meV, going from 60 K ($\approx 1/2\: T_C $) to 180 K
($\approx 3/2\:T_C$), which is in the same order of magnitude as
the 0.1 eV predicted by Furukawa. Qualitatively, on the other
hand, we find $\Phi_{\scriptstyle FM}>\Phi_{\scriptstyle PM}$,
which is exactly opposite to his conclusion (and to our intuitive
band picture of Fig.\  1).

\section{discussion}
A number of explanations can be put forward to explain this latter
disagreement. The most simple solution would be to infer from this
measurement that the tetragonal distortion is not large enough to
split the $e_g$ level into $x^2-y^2$ and $3z^2-r^2$ levels. In
that case, the $e_g$ majority spin band, which can contain a
maximum of two electrons, is only 30\% filled, and one would
indeed expect to find $\Phi_{\scriptstyle FM}$ larger than
$\Phi_{\scriptstyle PM}$. This however, would put us in
disagreement with the experiments by Park {\sl et al.}\\ To find a
general constraint for the solution to this problem, let us look
at it from the more general point of view of thermodynamics. In
the beginning of the nineties a universal relation was deduced for
the behaviour of the chemical potential at the transition
temperature of any second order phase transition\cite{marel}. It
states that there will be a change $\Delta$ in slope of $\mu$
versus temperature at $T_C$, provided that $T_C$ depends on
particle density $n$:
\begin{equation}
\Delta\left(\frac{d\mu_e}{dT}\right) = \Delta C \frac{d\ln
T_c}{dn_e}
\end{equation}
In which the subscript $e$ indicates that we are dealing with the
electronic part of both $\mu$ and $n$. Since this formula was
derived for constant pressure as well as for constant volume
conditions, the specific heat $C$ can be taken to be either $C_V$
or $C_p$ as long as the chemical potential measurement is
performed under the same conditions. We hereby define the
difference $\Delta$ (both in $C$ and in $d\mu/dT$) as the value
above $T_C$ (the disordered state) minus the value below $T_C$
(ordered state). In our experiments we have measured the work
function, which means that the change at $T_C$ will have the
opposite sign compared to $\mu$. From our data set we can only get
a rough estimate (assuming $d\mu/dT=0$ above $T_C$) of
$\Delta(d\mu/dT)=-\Delta(d\Phi/dT)\approx -4\cdot 10^{-4}$ eV/K.
The change in the specific heat was obtained from figure 2b in the
paper of Gordon {\sl et al.}\cite{gordon}. We find a change
$\Delta C = -54$ Jmol$^{-1}$K$^{-1}$ or $-8.0\cdot 10^{-5}$ eV/K
referred to a unit of MnO$_2$, since the chemical potential is
influenced (mainly) by the processes within the MnO$_2$ layers. To
find the slope of $T_C$ versus particle density for the double
layered manganites, we used the data taken from three different
papers\cite{kubota,ling,medarde} and averaging the slopes we found
for $x=0.4$ that $d\ln T_C/dx=-3.3 \pm 0.4$ (see Fig.\  4).
\begin{figure}[ht]
\centerline{\epsfig{figure=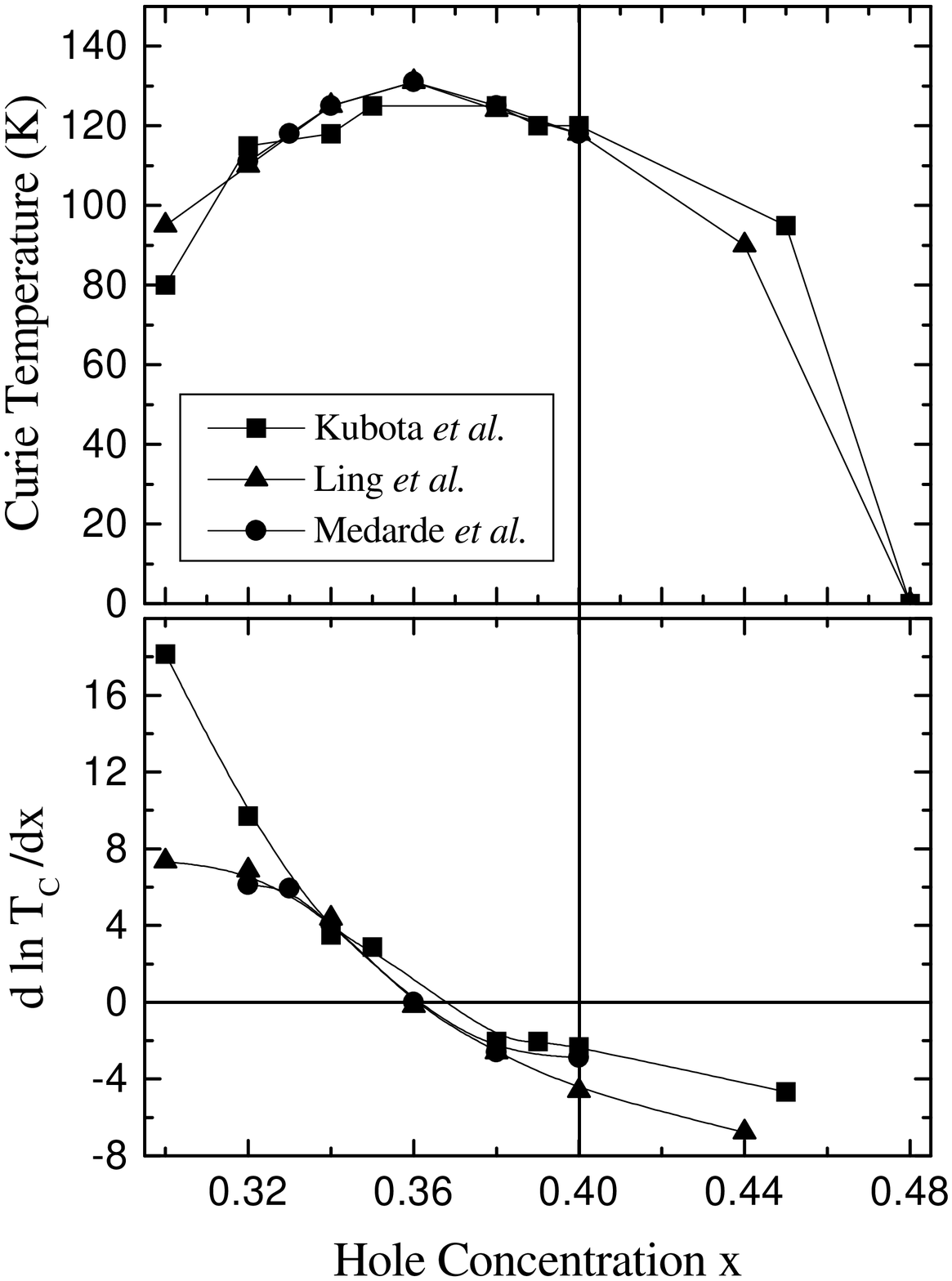,width=8.4cm,clip=}} \caption{Top
panel: Curie temperature as a function of hole doping for the
layered manganites taken from ref.\  15, 16 and 17 respectively.
Bottom panel: d ln T$_C$/dx versus hole doping.} \label{tc}
\end{figure}
This is, however, the slope depending on the concentration of
holes, for electrons we should therefore use the positive value.
Combining these findings in equation 2, we find that the right
hand side gives us $-3\cdot 10^{-4}$, which is in rather good
agreement with our value for the change in slope of $d\mu/dT$ of
$-4\cdot 10^{-4}$.\\ Another important fact we can obtain from
Fig.\  4, is that in all three data sets, the maximum $T_C$ occurs
around $x\simeq 0.36$. Below that, $d\ln T_C/dx$ is positive
(negative) for holes (electrons) and the other way around after
that. This is where the difference between the theoretical
prediction of the double-exchange model and our experimental
findings stem from. In the DE model, $T_C$ will increase from 0 to
$T_{M\!A\!X}$ in the range $x=0$ to $x=0.5$, after which it will
continuously decrease until it reaches $T_C$=0 at $x=1$ again. Our
sample has a hole concentration that lies exactly in a range
($0.36 < x < 0.5$) where the experimentally determined $T_C$
versus $x$ curve deviates from the theoretical DE prediction. We
therefore expect that a sample with a hole concentration $x$
smaller than 0.36 will show an {\sl increase} in $\mu$ in the
ferromagnetic phase, in accordance with Furukawa's prediction for
the DE model.\\ Before concluding, we would like to comment on the
fact that our sample is quasi two-dimensional, and that the
predictions made by Furukawa are based on a simple three
dimensional picture. The consequence of this is mainly that
short-range magnetic correlations can remain above
$T_C$\cite{moritomo} in bilayered samples like ours, almost up to
room temperature, and they tend to smear out the effect in $\mu$
at the phase transition over a larger temperature range. Since we
still observe a clear change comparing measurements below and
above $T_C$, we claim that these short range correlations are not
strong enough to obliterate the effects of the 3D long range order
setting in at $T_C$. Lastly, there has been some debate regarding
the precise nature of the phase transition in
\lsmo\cite{osborn,gordon}, but since no dependence on heating rate
was observed in the measurements of Gordon {\sl et al.} we believe
the phase transition to be at least of second order, which is an
important ingredient in our explanation.

\section{conclusion}
From photoemission measurements on slightly biased \lsmo we found
a decrease in the work function with increasing temperature below
$T_C$, and a roughly constant value above $T_C$. The quantitative
decrease is 55 $\pm$ 10 meV, going from 60 K to 180 K. Although
the number of measured temperatures is small, we are confident
that our measurements do show the true dependence on temperature,
based on a number of reasons: 1) All our successful cleaves
(always performed at 60 K) on \lsmo show the same downward trend
with temperature, irrespective of the low absolute value of the
work function at 60 K, 2) The presented measurement is
self-consistent within the two temperature cycles, 3) As a
function of increasing surface degradation the work function
increases, opposite to the temperature trend, 4) The measured
effect in \lsmo is far more substantial than the deviations with
temperature measured on a poly-crystalline silver sample.\\ Our
measurements show the opposite trend to that predicted by
Furukawa. We believe this stems from the fact that in the real
\lsmox system, the behaviour of $T_C$ with doping $x$ is different
than that inferred from a simple double-exchange picture. The
reason for this is that, in the DE model, the influence of the
intra-atomic exchange interaction $J_{e\!x}$, and the on site
Coulomb repulsion energy $U$, although initially taken into
account for selecting the dominant processes, are from there on
neglected. It is however obvious that, neither $U$ ($\approx$ 4
eV\cite{chainani} or 7 eV\cite{bocquet}) nor $J_{e\!x}$
($=3J_H\approx 2.7$ eV\cite{chpark}, in Mn$^{4+}$) can be
considered infinite with respect to the bandwidth W ($\approx
1-1.5$ eV). Therefore the continuing influence of $U$ and
$J-{e\!x}$ on the charge dynamics of the manganites is not to be
discarded. In strongly correlated electron models where these
parameters, together with hybridization, {\sl are} kept in play,
one generally finds asymmetrical doping behaviour of physical
properties, such as the spectral weight function in photoemission,
a direct link to the kinetic energy per particle\cite{eskes}.
Consequently, it is then no surprise that for the manganites the
transition temperature also does not peak at half-filling as the
DE model predicts, but rather at lower hole doping $x\approx 0.36$
(analogous to the perovskite manganites: $x\approx 0.31$, and the
high Tc cuprates: $x\approx 0.2$). In order to come to a full
understanding of the manganites, we want to emphasize therefore
that it is crucial to retain the electron correlation parameters
in the description of the charge dynamics and thus to go beyond
the effective single electron approximation of double-exchange.\\
Our findings are furthermore consistent with a general
thermodynamical relation for the difference in $\Phi$ across a
second-order phase transition. It validates the observed trend in
our measurements and links it to the difference between the simple
DE model and the real complexity of the manganites. It would be
interesting to further test this hypothesis by performing a work
function measurement on a \lsmox sample with $x<0.36$, where the
slope of d$T_C$/dx is opposite to ours and agreement with the DE
model would be expected. Our quantitative value for the work
function change is in accordance with the order of magnitude
inferred by the double-exchange model, suggesting $J_{e\!x}$ and
$U$ are large enough, compared to the bandwidth W, for
double-exchange to be the right starting point. Most importantly,
the observed chemical potential difference is large enough for the
explanation put forward by Klein {\sl et al.}, regarding the way
grain boundaries affect (magneto)-resistive properties in films,
to be valid.
\\
\begin{center}
{\bf ACKNOWLEDGEMENTS}
\end{center}
We would like to thank N. Furukawa, I. S. Elfimov, D.I. Khomskii,
D. van der Marel and M. Velazquez for useful discussions and their
valuable contributions along the way. This research was supported
by the Netherlands Foundation for Fundamental Research on Matter
(FOM) with financial support from the Netherlands Organization for
the Advancement of Pure Research (NWO). The research of MAJ was
supported through a grant from the Oxsen Network an the research
of LHT has been made possible by financial support from the Royal
Dutch Academy of Arts and Sciences.


\begin{references}
\bibitem{comment} We assume that the band that Furukawa uses for
his calculations can also contain 1 electron at maximum and in his
case then will be 80\% filled.
\bibitem{XPS} We have performed XPS measurements on freshly
cleaved and aged samples and observed a change in energy and a
splitting in both the lanthanum $3d$ and oxygen $1s$ peaks with
aging. The simplest explanation is assuming a different La-O ratio
at the surface in aged samples.
\end{references}
\end{document}